\newcommand{\be}{\begin{equation}}
\newcommand{\ee}{\end{equation}}
\newcommand{\bea}{\begin{eqnarray}}
\newcommand{\eea}{\end{eqnarray}}

\documentstyle[12pt,epsfig]{article}

\begin{document}
\renewcommand{\thefootnote}{\fnsymbol{footnote}}
\thispagestyle{empty}
\rightline{LNF-97/016 (P)}
\rightline{gr-qc/9704039 \vspace{1cm}}
\begin{center}
{\bf EFFECTIVE GRAVITY AND $OSp(N,4)$ INVARIANT MATTER} \vspace{1cm} \\
S. Bellucci\footnote{e-mail: bellucci@lnf.infn.it}
                    \vspace{0.5cm} \\
INFN-Laboratori Nazionali di Frascati, P.O.Box 13, 00044 Frascati,
            Italy
\vspace{2.5cm} \\
{\bf ABSTRACT} 
\end{center}
We re-examine the $OSp(N,4)$ invariant
interacting model of massless chiral and gauge superfields, whose
superconformal invariance was instrumental, both in proving
the all-order no-renormalization of
the mass and chiral self-interaction lagrangians, and in determining
the linear superfield renormalization needed. We show that
the renormalization of
the gravitational action modifies only the
cosmological term, without affecting higher-order tensors. This could
explain why the effect of the cosmological constant is shadowed by the effects
of newtonian gravity.
\vfill
\begin{center}
March 1997
\end{center}
\setcounter{page}0
\renewcommand{\thefootnote}{\arabic{footnote}}
\setcounter{footnote}0
\newpage
\section{Introduction}

The introduction of locally supersymmetric
theories [1] was motivated
by the wish for
a unified description, encompassing the theories of elementary
particles and gravity.
The resulting supergravity is not renormalizable, but
its large symmetry provides powerful constraints, such as the
vanishing of the lepton anomalous magnetic moment, as required
by the supersymmetry of the theory [2].
The finiteness of the fermionic (and therefore
also the bosonic) contribution to it, which can be traced back to an effective
chiral symmetry in the gravitino sector [3], has been
checked in [4,3] and, more recently in [5], making use of supersymmetry
preserving regularization schemes.

An introduction to the effective action in quantum gravity can be found
in [6]. The quantum corrections to the low-energy
limit of a theory coupling gravity to scalar fields have been
computed [7]. The equivalence principle has been invoked,
in order to reduce the terms with
an arbitrary number of derivatives in the effective theory [8].
This principle also constrains the spin-1 and spin-0
partners of the graviton in the $N=2,8$ supergravity multiplets [9,10].

$OSp(N,4)$ invariant models in the fixed four-dimensional
background of anti-de Sitter space ($AdS_4$)
occur as both ground state solutions of gauged extended supergravity
theories (see [11] and references therein) and
vacuum configurations for superstrings [12]. 
The $OSp(N,4)$ invariant generalization of the Wess-Zumino model [13]
is the simplest one. We recall that
the Wess-Zumino model with softly broken supersymmetry in de Sitter space
plays a role in the Affleck-Dine mechanism [14] for baryogenesis, in contrast
to the maximal symmetry of $AdS_4$ which grants the existence of global
supersymmetry. This mechanism is effective for supersymmetric grand unified
theories, as the quantum corrections do not affect the flat directions
in the superpotential, owing to the no-renormalization theorem [15].

Very recently, the one-loop effective potential along a flat direction in
this model has been calculated [16]. In [17] the one-loop effective action
was
computed for scalar-QED, taking into account the large-scale configurations
that change the topology. Also, the question of the dynamics of a superstring
propagating in $AdS_4$, with the $OSp(1,4)$ supersymmetry group, deserves
further study, within a geometrical framework, especially in connection with
the underlying algebraic structure of the $W$-algebra extension of
two-dimensional conformal symmetry (see, e.g. [18-21]).

The renormalization procedure for $OSp(N,4)$ invariant theories breaks
the naturality implied by the no-renormalization theorem and so allows
{\it all} classically invariant counterterms to appear in the
divergent structure of the quantum effective action [22-34]. This breakdown 
of the no-renormalization theorem in $AdS_4$ forces us to
introduce a linear superfield in the effective action, for the purpose
of renormalization. However, the corresponding modification of the classical
potential does not induce the breaking of supersymmetry invariance [25,11].
The superconformal invariance of the model with interacting chiral and
real gauge superfields in $AdS_4$, following the line suggested in Ref.
[35], allowed us
to prove to all orders in the perturbative series the non renormalization
of the mass nor the cubic interaction action [34].

We organize the present work as follows.
We begin in sec. 2 with recalling the superfield formulation of
the $AdS_4$ interacting model of chiral and gauge
superfields.
The superconformal invariance of the massless model allows us to
implement an expansion in the curvature effects, in terms of the interaction
vertices of the quantum model.
In sec. 3 we present the renormalization of the
gravitational action, based on the use of $OSp(N,4)$ superfield techniques,
and propose some interpretation for the shadowing of the effect of the
cosmological constant by the effects of newtonian gravity.
We draw our conclusions in sec. 4.

\section{Interacting chiral and real gauge superfields}

For the purpose of fixing notations, we briefly recall in this section a
superspace approach [35,33,34] to the $OSp(N,4)$ invariant
theory of a supergravity multiplet coupled to 
interacting chiral and real gauge supermultiplets.

In order to choose $AdS_4$ as a background space for the matter
and gauge model,
we set the supergravity prepotential $H$ to zero and introduce the
background only through the
compensator $\phi$. Its equation of motion
\begin{equation}
{\overline D}^2{\overline\phi}=\alpha\phi^2
\end{equation}
can be obtained from the
action for supergravity with the cosmological term
\begin{equation}
S=-\frac{3}{\kappa^2}\int d^4xd^4\theta E^{-1} +(
\alpha\frac{1}{\kappa^2}\int d^4xd^2\theta \phi^3 +h.c.) \quad ,
\end{equation}
yielding the solution with a regular behaviour at infinity
\begin{equation}
\phi =\frac{1}{1-\alpha{\overline{\alpha}}x^2/4}-
\frac{{\overline{\alpha}}\theta^2}{(1-\alpha{\overline{\alpha}}x^2/4)^2}
\quad ,
\end{equation}
with the inverse determinant $E^{-1}={\overline{\phi}}\phi$.
Then, by applying this solution to the construction of invariant
actions
of the general and chiral type, we can formulate different
supersymmetric matter models in the given background. We recall
the expression of the covariant derivatives in terms of the
$\phi$ field
\begin{equation}
{\overline{\nabla}}_{{\dot{\alpha}}}=\phi^{-1}{\overline{\phi}}^{1/2}
{\overline{D}}_{{\dot{\alpha}}}\quad ,\quad
{\nabla}_{\alpha}={\overline{\phi}}^{-1}\phi^{1/2}D_{\alpha}\quad .
\end{equation}

The theory of interacting chiral ($\eta$) and real gauge superfields
is described (in the gauge-chiral representation) by the
action [35]
\begin{eqnarray}
S(\eta ,{\overline{\eta}},V)
&=&\int d^4xd^4\theta E^{-1} {\overline{\eta}}{}_j
[exp(V)]{}^j{}_i\eta^i\nonumber \\
&+&[\int d^4xd^2\theta \phi^3 W^2
+\int d^4xd^2\theta \phi^3 (m\frac{1}{2}\eta^2+
\lambda \frac{1}{6}\eta^3)+h.c.] \quad ,
\end{eqnarray}
with $V^i_j=V^A(T_A){}^i{}_j$, and where $(T_A){}^i{}_j$ is a matrix
representation
of the generators of the gauge group that leaves this action invariant.
This model possesses a partial superconformal invariance, which
has been exploited [33,34], in order to treat perturbatively the
effects of the background curvature, when carrying out the
renormalization procedure that yields its quantum effective action.
All $\phi$-dependence in the free-field functional
integral can be removed by carrying out a superconformal
transformation, in accordance with the canonical weights
of the matter and gauge fields and their sources
\begin{equation}
{\hat{\eta}}=\phi\eta\quad ,\quad
{\hat{W}}{}_\alpha=\phi^{3/2}W_\alpha\quad ,\quad
\hat{J}=\phi^2 J\quad ,\quad
\hat{J}{}_V=\phi
{\overline{\phi}} J_V
\quad .
\end{equation}
Hence the quantum model can be described in the
most natural form in terms of the transformed fields,
i.e. defining the effective action in terms of the hatted
fields.

The definition of ${\hat{W}}$ in terms of the familiar derivatives
in flat background reads
\begin{equation}
{\hat{W}}{}_\alpha=i{\overline{D}}{}^2 D_\alpha V\quad .
\end{equation}
The covariantization of this expression in the Yang-Mills chiral representation
\begin{equation}
{{W}}{}_\alpha=i({\overline{\nabla}}{}^2+\alpha)
exp(-V){\nabla}_{\alpha} exp(V)
\end{equation}
gives $W_{\alpha}$ in terms of the background covariant derivatives
(2.4).
After the superconformal transformation, the gauge-fixing procedure can
be carried out, along the line of the flat background theory. The resulting
gauge propagator reads (in the Fermi-Feynman gauge, with $\xi =1$)
\begin{equation}
<VV>{}_0=-\frac{1}{p^2}\delta^4 (\theta -\theta ') \quad .
\end{equation}
It is worthwhile
to notice that, as a consequence of superconformal invariance, the
ghost propagators and vertices of the flat space-time theory, along
with the usual flat space-time $D$-algebra, remain
intact in $AdS_4$.

Enforcing the boundary conditions
needed, in order to preserve supersymmetry for the scalar and spinor
propagators in $AdS_4$ [32], and evaluating the free-field functional
integral, yields the vacuum expectation values [33]
\begin{equation}
<T\hat{\eta}(x')\hat{\overline{\eta}}(0)>=\frac{1}{4\pi^2}
\hat{\overline{D}}{}^2 D^2\delta^4(\theta-\theta ')
\frac{1}{(x')^2} \quad ,\quad
<T\hat{\eta}(x')\hat{\eta}(0)>=\frac{1}{16\pi^2}
(|\alpha|^2+2|\alpha|^3\theta '\theta )
\quad .
\end{equation}
From the generating functional one can read the vertex contributions
involving the field $\hat{\eta}$. It turns out [33,34] that
there are no $\phi$'s at every such vertex (in $D=4$),
with the only exception of a vertex quadratic in $\hat{\eta}$,
which appears with a factor $\phi$.
The conclusion is that, for our purposes, we can handle
the quantum system of $N=1$
super Yang-Mills coupled to matter scalar superfields in $AdS_4$ in
a way similar to the corresponding theory in a flat background, with
the only difference of including in the Feynman rules
the additional quadratic vertex
\begin{equation}
\frac{1}{2}m\phi+h.c.
\quad .
\end{equation}
The residual explicit $\phi$ dependence of the latter
reflects the deviation of the theory from a
superconformal one, owing to the introduction of a mass term.
A remarkable feature of the above rescaling, which effectively
removes $\phi$ from the superconformal invariant part of the
superfield action (with the caveat of possible anomalous contributions
[33]), is that it takes automatically into consideration the need to
resort to some
perturbative approach in the effects of the curvature of the
background space, leading us naturally to introduce the above implicit
expansion in the compensator $\phi$.

\setcounter{equation}0
\section{The renormalization of the gravitational action}

Herewith we describe our main result, in an attempt to
explain the shadowing of
the cosmological constant.
We build upon our previous work [33,34] and carry out
the renormalization of the gravitational
action
induced by $OSp(N,4)$ invariant matter multiplets in curved space.
Here great care is needed, as every sign at each step is crucial.

We start
by discussing the renormalization of the gravitational action induced by
a matter chiral superfield, through the presence of the Feynman diagram
in figure 1. This yields the following divergent contribution:
\begin{equation}
am^2\frac{1}{\epsilon}\int d^4 xd^4 \theta \phi{\overline{\phi}}
\quad ,\quad 
\end{equation}
with $a>0$ in any case. One can doubt about the fact of interpreting this
diagram as a renormalization of the pure supergravity action
\begin{equation}
-\frac{3}{\kappa^2}\int d^4 xd^4 \theta \phi{\overline{\phi}}
\quad ,\quad 
\end{equation}
or the cosmological superspace action
\begin{equation}
\frac{\alpha}{\kappa^2}\int d^4 xd^2 \theta \phi^3+h.c.
\quad .\quad 
\end{equation}
There are clear reasons to support this point of view,
which goes in the direction of considering the above contribution as a
renormalization of the second term, i.e. the cosmological action. In
order to pursue this idea, we can translate the result in components,
to read as follows:
\begin{equation}
am^2\alpha_R^2\frac{1}{\epsilon}\int d^4 x \sqrt{-g}
\quad .\quad 
\end{equation}
Here, and throughout this section, we denote with subscript indices $R,B$
the renormalized and the bare parameters of the action, respectively..

\begin{figure}[t]
\centerline{\epsfig{file=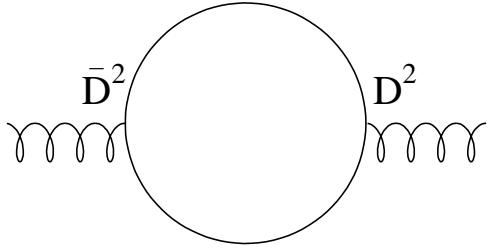,height=8cm}}
\caption{One-loop renormalization of gravitational actions from chiral matter
(the ${\hat{\eta}}$ solid lines):
the curly lines denote the compensator $\phi$ and its h.c.
${\overline{\phi}}$.} 
 \protect\label{fig1}
\end{figure}

On the other hand, we know that the final expression for the gravitational
actions has to be
\begin{equation}
S=-\frac{1}{\kappa^2}\int d^4 x \sqrt{-g}R(r) +
\frac{6}{\kappa^2}\int d^4 x \sqrt{-g}\alpha_R^2
\quad ,\quad 
\end{equation}
where, let us say, $R$ is restricted to metrics of the anti-de Sitter type,
with arbitrary radius $r$. Then it is clear to us that the contribution
(3.4) can be consistently interpreted as a renormalization of only the
cosmological term, i.e. the second term on the r.h.s. of Eq. (3.5). This
interpretation is also compatible with the fact that the renormalization of
the higher order gravitational tensors (i.e. the contributions of order
$\alpha_R^4$) does not take place. In our opinion, this is not a coincidence,
rather it means that any vacuum diagram can be seen as a renormalization of
only one parameter in the gravitational action, namely $\alpha_R$.

So, let us write then
\begin{equation}
am^2\frac{1}{\epsilon}\int d^4 xd^4 \theta \phi{\overline{\phi}}=
am^2\frac{1}{\epsilon}\alpha_R\int d^4 xd^2 \theta \phi^3
\quad ,\quad 
\end{equation}
where we make use of the equation of motion (2.1) for the chiral compensator
superfield $\phi$. Considering this contribution alone, we would have
\begin{equation}
\alpha_B=\alpha_R-
am^2\alpha_R\frac{1}{\epsilon}\kappa^2\equiv Z_{\alpha}\alpha_R
\quad ,\quad 
\end{equation}
but then it is obvious that this renormalization factor $Z_\alpha$ cannot
lead to a running $\alpha_R$, since the above $m$ and $\kappa$ parameters
do not depend on the renormalization scale $\mu$ (the $m$ parameter that
appears here cannot be anything, other than the bare mass $m_B$).

\begin{figure}[t]
\centerline{\epsfig{file=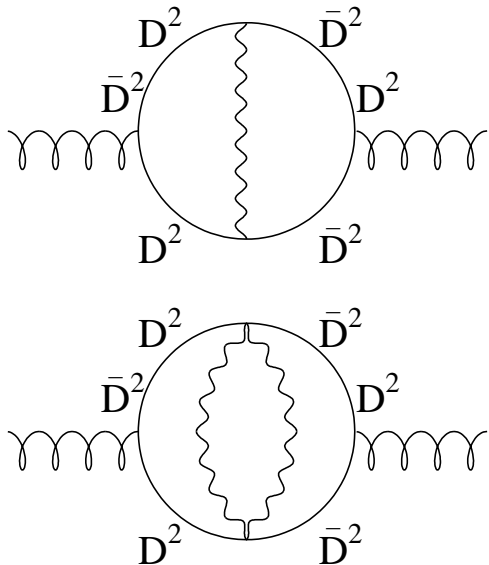,height=8cm}}
\caption{$O(g^2)$ two-loop corrections to the cosmological
renormalization factor $Z_{\alpha}$ in (3.7);
the wavy lines denote the gauge superfields.}
 \protect\label{fig2}
\end{figure}

In this way, we reach the conclusion that the relevant term in $Z_\alpha$
is the two-loop contribution to the vacuum, what forces us to introduce
gauge interactions in the game. In fact, when introducing gauge
interactions, of all the plethora of diagrams that one can imagine to the
order $g^2$ in the gauge coupling constant, we believe there are only two
that survive, after performing the $D$-algebra of the covariant derivatives,
i.e. the diagrams in figure 2. We give, in the following, the computation
of the first graph (at the top of figure 2),
as the other one (at the bottom of figure 2)
looks really frightening to compute,
and anyhow it cannot change the conclusion of this story. Using integration
by parts for the superspace covariant derivatives and discarding a finite
remainder given in figure 3, one can identify the divergent parts of
the first graph in figure 2 and the diagram in figure 4.

\begin{figure}[t]
\centerline{\epsfig{file=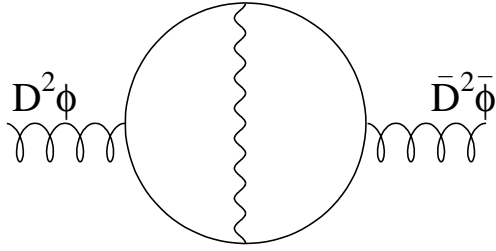,height=8cm}}
\caption{A finite contribution to the gravitational action produced, when
integrating by parts the superspace covariant derivatives in the Feynman
diagram at the top of figure 2, in order to obtain the graph of figure 4.}
 \protect\label{fig3}
\end{figure}

The diagram in figure 4 yields the amplitude ${\cal V}$
\begin{equation}
{\cal V}=
m^2g^2
\int\frac{d^D k}{(2\pi)^D}\int\frac{d^D q}{(2\pi)^D}
\frac{1}{k^2}\frac{1}{q^2}
\frac{1}{(k-p)^2}\frac{1}{(k+q)^2}
\quad ,\quad 
\end{equation}
where we work in dimension $D=4-\epsilon$. The first integral
can be carried out, yielding the amplitude
\begin{equation}
{\cal V}=
m^2g^2\frac{1}{(2\pi)^{4-\epsilon}}
\pi^{2-\epsilon /2} \Gamma(\frac{\epsilon}{2})
B(1-\frac{\epsilon}{2},1-\frac{\epsilon}{2}){\cal I}
\quad ,\quad 
\end{equation}
where
\begin{equation}
B(1-z,1-w) \equiv \int_0^1 dy\frac{1}{y^z(1-y)^w}
\end{equation}
and we define the momentum integral
\begin{equation}
{\cal I}\equiv
\int\frac{d^D k}{(2\pi)^D}
\frac{1}{k^2}\frac{1}{(k-p)^2}\frac{1}{(k^2)^{\epsilon /2}}
\quad .\quad 
\end{equation}

\begin{figure}[t]
\centerline{\epsfig{file=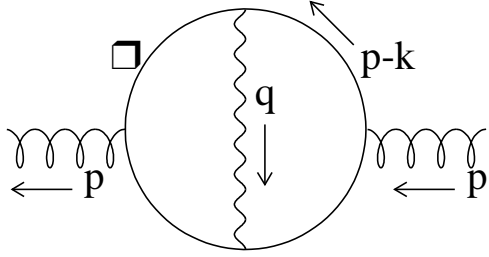,height=8cm}}
\caption{$O(g^2)$ two-loop Feynman diagram corresponding to the amplitude
${\cal V}$ in (3.8); the two independent momenta (loop variables)
of the internal lines are
explicitly indicated, together with the momenta of the external
compensators.}
 \protect\label{fig4}
\end{figure}

Also the integral ${\cal I}$ can be evaluated. Hence, using
standard properties of the $B$-function, we get the result
\begin{equation}
{\cal V}=
m^2g_R^2\frac{1}{(2\pi)^{8-2\epsilon}}
\pi^{4-\epsilon}\frac{2}{\epsilon}\Gamma(\epsilon)\frac{1}{1-\epsilon}
[\Gamma(1-\frac{\epsilon}{2})]^3
\frac{1}{\Gamma(2-3\epsilon /2)}
\Biggr( \frac{\mu^2}{p^2}{\Biggr)}^{\epsilon}
\quad .\quad 
\end{equation}
Here we introduced the expression of the renormalized gauge coupling
constant $g_R$ in terms of the renormalization scale
\begin{equation}
g_R=g\mu^{-\epsilon}
\quad .\quad 
\end{equation}

Two remarks are in order about this result. First of all, the leading
divergence is of order $\epsilon^{-2}$, what makes it relevant for the
purpose of running $\alpha_R$. Secondly, it contains subleading nonlocal
divergencies of the form ${\mbox{log}}(p/\mu)$. We will have to prove that
they cancel with similar terms coming from the other graph in figure 2,
for this whole thing to make sense.
What is important about this other graph, apart from the cancellation of
the nonlocal divergencies, is that the leading divergence comes with
a positive sign (as it is the case for the amplitude ${\cal V}$
computed above). So let us assume that the whole contribution from figure
2 is of the form
\begin{equation}
bm^2g_R^2\alpha_R(\frac{1}{\epsilon})^2\int d^4 xd^2 \theta \phi^3
\quad ,\quad 
\end{equation}
{\underline{with $b>0$}}, plus perhaps $1/\epsilon$ subleading
divergencies that are not relevant, when studying the renormalization
group equation for $\alpha_R$ to the order $g_R^2$. Then, we have that
\begin{eqnarray}
Z_{\alpha} &=& 1-am^2\kappa^2\frac{1}{\epsilon}-bm^2\kappa^2g_R^2
(\frac{1}{\epsilon})^2\\
\frac{\mbox{d}\alpha_B}{\mbox{d}\mu} &=& 0=Z_\alpha
\frac{\mbox{d}\alpha_R}{\mbox{d}\mu}+\frac{\mbox{d}Z_\alpha}{\mbox{d}\mu}
\alpha_R\\
\frac{\mbox{d}\alpha_R}{\mbox{d}\mu} &=& -\frac{1}{Z_\alpha}
\frac{\mbox{d}Z_\alpha}{\mbox{d}\mu}\alpha_R\nonumber \\
&\approx & -\frac{1}{1-am^2\kappa^2/\epsilon}[-2bm^2\kappa^2g_R
\frac{\mbox{d}g_R}{\mbox{d}\mu}(\frac{1}{\epsilon})^2]\alpha_R+O(g_R^3)
\nonumber \\
& = & -\frac{1}{1-am^2\kappa^2/\epsilon} (bm^2\kappa^2g_R^2
\frac{1}{\mu}\frac{1}{\epsilon})\alpha_R+O(g_R^3)
\quad .\quad 
\end{eqnarray}

In the limit $\epsilon\to 0$, we have then
\begin{equation}
\frac{\mbox{d}\alpha_R}{\mbox{d}\mu} = \frac{b}{a}g_R^2\frac{1}{\mu}\alpha_R
\quad ,\quad 
\end{equation}
so that one can easily guess the kind of theories, in which the effective
cosmological constant goes to zero in the infrared, as a power of $\mu$.
We can then write that, if $g_R^2=$ constant
\begin{equation}
\alpha_R=\alpha_0\Biggr( \frac{\mu}{\mu_0}{\Biggr)}^{bg_R^2/a}
\quad .\quad 
\end{equation}
The important thing, which we checked repeatedly, is that the exponent is
{\it positive}, though it is more obscure, under which circumstances it could
be bigger than one.

\section{Conclusion}

We cannot add much more to the above considerations, apart from the fact that,
if the interpretation $\mu^2\approx R$ in a gravitational measurement is
plausible, then this could explain why the effect of the cosmological
constant is always shadowed, no matter what the value of $\alpha_0$ might
be, by the effects of newtonian gravity.

The calculation of the stress-tensor anomaly in $AdS_4$ supersymmetry
showed [36] that the choice of the vacuum, around which
the model is quantized, does not affect the renormalization of the 
purely geometrical tensors in the effective action, nor the trace anomaly
induced by matter multiplets invariant under the supersymmetry
group $OSp(N,4)$ in curved space.
These quantities are independent also from the boundary conditions for the
free-field propagators, as proven in Ref. [36].
We remark, in passing, that the conformal anomaly, as an integrability
condition
for the supersymmetric sigma models corresponding to superstring theories,
was obtained in Refs. [37-41].

Next, we wish to compare the work of Elizalde
and Odintsov (E-O) [42] with
our result (3.19) on the running and the consequent exponential shadowing
of the cosmological constant.
The two results appear to be similar, with ours
as a particular case, at least at first sight. Notice however that our
paper is of wider interest
in what concerns keeping a global (anti-de Sitter) supersymmetry
of the curved background space. Indeed our work and [42] may be
considered as complementary, in many respects.

The paper by E-O contains a phenomenological analysis of theories
that are finite in flat spacetime. A class of such theories is considered in
[42] interacting with an external gravitational field (including a
nonminimal term linear in the curvature and quadratic in the massless
scalar matter field).
Our calculation applies specifically to the (globally) supersymmetric
background (ground state) solution for SUSY GUTS, extended SUGRA,
superconformal invariant theories (superstrings). As in our calculation
the (global) supersymmetry of this ground state solution is maintained,
consequently the contribution of the supersymmetric particles is determined
explicitly.

In addition, the role of a nonvanishing mass for the matter
fields is perfectly clear and is included in our analysis (E-O only
consider a massless theory).
Finally, in E-O a fine tuning is needed, in order to avoid in their
Eq. (19) the unrealistic growth of Newton's constant, and/or to obtain
the screening of the cosmological constant. We obtain such screening
for $\alpha_R$ directly and without fine tuning.

We stress that we consider the superconformal invariant theories to be
minimally interacting with the external gravitational background.
Our main motivation is to keep the global supersymmetry of the
background, which requires referring to the minimal n=-1/3 compensator
$\phi$. It was shown long ago in ref. [35] (page 336 section 5.7)
that nonminimal n implies spontaneous breakdown of N=1 supersymmetry.

In our paper we have recalled several developments that simplified
the procedure in practical calculations for the renormalization
of our class of theories. In particular the superconformal rescaling
is an improvement that can be useful in future applications. In this
respect we hope that the discussion of the method given in this work
will find further use.

\section*{Acknowledgement}

We thank with great pleasure Jos\'e Gonz\'alez, for participating to the
early stages of this research. We thank the referee of this paper for
calling Ref. [42] to our attention.


\end{document}